\documentclass[fleqn,twoside]{article}
\usepackage{espcrc2}
\usepackage{amsmath,amssymb}

\usepackage{graphicx}

\newcommand{\be}{\begin{equation}}
\newcommand{\ee}{\end{equation}}
\newcommand{\bea}{\begin{eqnarray}}
\newcommand{\eea}{\end{eqnarray}}

\title{Neutrino oscillograms of the Earth and
CP violation in neutrino oscillations}

\author{E.~Kh.~Akhmedov\address{Max Planck Institute for Nuclear Physics, 
        Saupfercheckweg 1,\\ 69117 Heidelberg, Germany}} %
       
\begin{document}

\begin{abstract}
An analysis of 3-flavour neutrinos oscillations inside the Earth is 
presented in terms of the oscillograms -- contour plots of oscillation 
probabilities in the plane neutrino energy -- nadir angle. Special 
attention is paid to CP violation in neutrino oscillations in the Earth. 

\vspace{1pc}
\end{abstract}

\maketitle

\section{INTRODUCTION}
Neutrino oscillograms of the Earth are contours of constant oscillation
or survival probabilities in the plane of neutrino energy and nadir 
or zenith angle of neutrino trajectory. They encode information on both the 
neutrino parameters and the Earth density profile and proved to be a very 
useful and  illuminating tool for analyzing neutrino oscillations in the 
Earth.

The oscillograms exhibit a very rich structure with local and global
maxima and minima (including the MSW resonance maxima in the mantle and
core of the Earth), saddle points and the parametric enhancement ridges  
(Fig.~\ref{fig:1}). It was shown in \cite{Akhmedov:2006hb} that all these 
features can be readily understood in terms of various realizations of 
just two conditions: the generalized amplitude (collinearity) and phase 
conditions. The study in \cite{Akhmedov:2006hb} was performed in the 
limit $\Delta m_{21}^2=0$; in the present talk, based on 
\cite{Akhmedov:2008qt}, the results for $\Delta m_{21}^2\ne 0$ are 
presented. 

\section{OSCILLOGRAMS FOR 3-FLAVOUR NEUTRINO OSCILLATIONS}
\begin{figure}[htb]
\hspace*{0.6cm}
{\includegraphics[width=5.4cm,angle=90]{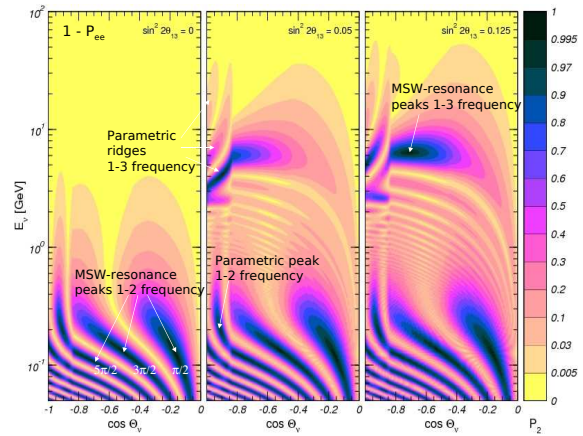}}
\vspace*{-1.0cm}
\caption{\small Oscillograms for $1-P_{ee}$ for three different values 
of $\theta_{13}$.}
\label{fig:1}
\end{figure}

In the full 3-flavour framework with $\Delta m_{21}^2\ne0$ 
and $\theta_{13}\ne 0$, the oscillograms have non-trivial structure 
both at low and high energies. Oscillations at low energies are 
dominated by the solar channel, whereas those at higher energies are
mainly driven by the atmospheric parameters $(\Delta 
m_{31}^2,\theta_{13})$ (see Fig.\ref{fig:1}). 

\begin{figure}[htb]
\hspace*{-0.3cm}
  \includegraphics[width=7.6cm]{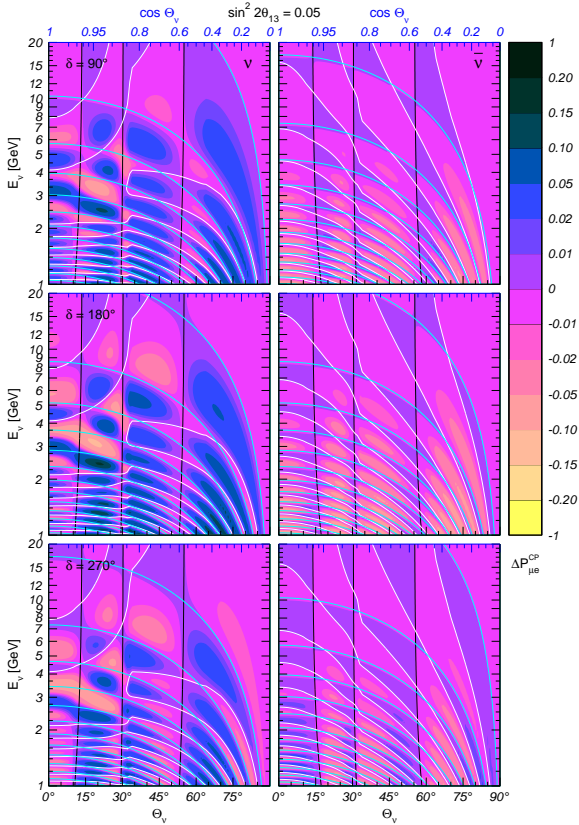}
\vspace*{-0.7cm}
  \caption{
    Oscillograms for the difference of probabilities $\Delta P_{\mu
    e}^{\rm CP} = P_{\mu e}(\delta) - P_{\mu e}(\delta_{\rm th})$
    with $\delta_{\rm th} = 0^\circ$. Shown are the solar (black),
    atmospheric (white) and interference phase condition (cyan) 
    curves.}
\label{fig:2}
\end{figure}

The main qualitatively new feature of 3-flavour oscillations as compared 
to 2-flavour ones is the possibility of CP violation. Its effect can be 
conveniently studied in terms of CP-oscillograms, which are defined as 
contour plots of equal probability difference 
\be
\Delta P_{\mu e}^{\rm CP} \equiv 
P_{\mu e}(\delta) - P_{\mu e}(\delta_{\rm th})\,,
\label{eq:1}
\ee
where $\delta$ is the true value of the CP violating phase and 
$\delta_{\rm th}$ is the assumed, or theoretical, value which we want to test. 
The CP-oscillograms have a rather complex domain-like structure (see 
Fig.~\ref{fig:2}); however, this structure can be readily understood and 
interpreted in terms of three grids of curves, as we discuss next. 
\vspace*{-0.2cm}

The oscillation probability $P_{\mu e}$ can be conveniently written as 
\be
P_{\mu e}=|c_{23}\, A_{e2}\, e^{i \delta}+s_{23}\,A_{e3}|^2\,,
\label{eq:2}
\ee
where the amplitudes $A_{e2}$ and $A_{e3}$ are independent of $\theta_{23}$ 
and CP-violating phase $\delta$. To leading order in small parameters $s_{13}$ 
and $\Delta m_{21}^2/\Delta m_{31}^2$, $A_{e2}$ depends only on the 
solar oscillation parameters and $A_{e3}$, only on the atmospheric ones: 
\be
A_{e2}\simeq A_S(\theta_{12},\,\Delta m_{21}^2)\,,~~~
A_{e3}\simeq A_A(\theta_{13},\,\Delta m_{31}^2)\,, 
\label{eq:3}
\ee
i.e., these amplitudes approximately coincide with, respectively, the 
solar 
and atmospheric contributions to the amplitude of $\nu_\mu\leftrightarrow 
\nu_e$ oscillations. {}From eq.~(\ref{eq:2}) we find
\bea
P_{\mu e}\!\!\!\! &=&\!\!\!\!
c_{23}^2 |A_S|^2 + s_{23}^2 |A_A|^2 ~~~~~\nonumber \\
&&\!\!\!\!+ 2 s_{23} c_{23} |A_S| |A_A| \cos(\phi + \delta)\,,
\label{eq:4}
\eea
where $\phi = arg(A_S A_A^*)$. Only the last term in (\ref{eq:4}) depends 
on $\delta$; the condition of vanishing $\Delta P_{\mu e}$ (i.e. 
$P_{\mu e}(\delta)=P_{\mu e}(\delta_{\rm th})$) therefore corresponds to
\be
|A_S| |A_A| \cos(\phi + \delta)=|A_S| |A_A| \cos(\phi + \delta_{th})\,.
\label{eq:5}
\ee
There are three non-trivial realisations of this condition:
\begin{itemize}
\item 
$A_S~=~0$ ~($\Rightarrow$ solar ``magic'' lines)

\item 
$A_A~=~0$ ~($\Rightarrow$ atm. ``magic'' curves)

\item 
$(\phi + \delta_{th}) = - (\phi + \delta) +
2\pi l$, ~~or~~\\
$
\phi = -(\delta + \delta_{th})/2 + \pi k.
$
\end{itemize}
There is also, of course, the trivial realization, $\delta_{\rm th}=\delta+ 
2\pi n$, when the true and assumed values of ~$\delta$ coincide.
The above three conditions give rise to three grids of curves, which delineate 
the different domains in the CP oscillograms (Fig.~\ref{fig:2}); they have a 
very simple physical interpretation. We shall now discuss these conditions 
in the energy region between the 1-2 and 1-3 MSW resonances; the results 
for other energy domains can be found in \cite{Akhmedov:2008qt}.

\subsection{Solar ``magic'' curves}
The condition of vanishing solar channel contribution to  the oscillation 
probability, $A_S=0$, in the constant matter density approximation can be 
written as $\sin^2(\omega_{12} L)=0$ (i.e. $\omega_{12} L=\pi n$), where 
$\omega_{12}$ is the solar oscillation frequency. At energies exceeding 
the 1-2 resonance energies in the mantle and in the core of the Earth, 
$E\gtrsim 0.5$ GeV, one has $\omega_{12}\simeq V/2$, where $V=\sqrt{2} G_F 
N_e$ is the matter-induced potential of neutrinos. 
The condition $A_S=0$ therefore takes the form
\be
    L(\Theta_\nu)  \simeq \frac{2\pi n}{V} \,,
\label{eq:6}
\ee
where $L(\Theta_\nu)$ is the nadir angle dependent length of the neutrino 
trajectory in the Earth.  Note that the condition (\ref{eq:6}) is energy 
independent and determines the baselines for which the ``solar'' contribution 
to the probability vanishes.  In the plane $(\Theta_\nu, E_\nu)$ it represents 
nearly vertical lines $\Theta_\nu \approx \text{const}$ 
(black lines in Fig.~\ref{fig:2}). 

There are three solar magic lines which correspond to $n = 1$ (in the
mantle domain) and $n = 2, 3$ (in the core domain). The existence of a
baseline ($L\approx 7600$ km) for which the probability of
$\nu_e\leftrightarrow \nu_\mu$ oscillations in the Earth is
approximately independent of the ``solar'' parameters ($\Delta m_{21}^2$,
$\theta_{12}$) and of the CP-phase $\delta$ was first pointed out     
in~\cite{Barger:2001yr} and later discussed in a number of publications 
(see~\cite{Huber:2002uy,Smirnov:2006sm} and references therein). 
This baseline 
was dubbed ``magic'' in~\cite{Huber:2002uy}.  The interpretation of this 
baseline as corresponding to vanishing ``solar'' amplitude $A_{e2}$, according 
to eq.~(\ref{eq:6}) with $n=1$, was given in~\cite{Smirnov:2006sm}. 
It was also shown there that for neutrino trajectories crossing the core of 
the Earth there exist two more solar ``magic'' baselines, corresponding to the 
oscillation phase equal $\pi n $ with $n=2$ and 3, and the existence of the 
atmospheric ``magic'' curves was pointed out.

\subsection{Atmospheric ``magic curves''}

The second realization of the condition (\ref{eq:5}) corresponds to 
vanishing atmospheric amplitude, $A_A=0$. In this case, like on the 
solar ``magic'' lines, the probabilities of $\nu_e\leftrightarrow \nu_\mu$ 
(as well as $\nu_e\leftrightarrow \nu_\tau$) oscillations are independent of 
the CP-violating phase. In addition, they do not depend on the 
atmospheric parameters $\Delta m_{31}^2$ and $\theta_{13}$. 

The properties of atmospheric ``magic'' curves can be easily understood in
the constant density approximation, in which the condition $A_A=0$ is 
satisfied when $\sin^2(\omega_{31} L(\Theta_\nu))=0$, i.e. $\omega_{31} 
L(\Theta_\nu)=\pi k$, $k = 1, 2, \dots$. For energies which are not too close 
to the atmospheric MSW resonance energy, this condition reduces to 
\be
E \simeq \frac{\Delta m_{31}^2 L(\Theta_\nu)}
{|4\pi k \pm 2 V L(\Theta_\nu) |} \,,
\label{eq:7}
\ee
which corresponds to the bent curves in the $(\Theta_\nu, E_\nu)$
plane (shown in white in Fig.~\ref{fig:2}). 

\subsection{The interference phase condition}

The third of the above mentioned realizations of the condition (\ref{eq:6}) 
depends on the interference phase between the amplitudes $A_S$ and $A_A$ 
and therefore we shall call it the interference phase condition. 
In the energy region between the 1-2 and 1-3 resonances we have $\phi \approx 
\Delta m_{31}^2 L/4E$, i.e. in the first approximation $\phi$ does not depend 
on the matter density. The interference phase condition then takes the form 
$\Delta m_{31}^2 L/4E=-(\delta + \delta_{\rm th})/2+\pi l$, or
\be
    E_\nu = \frac{\Delta m_{31}^2 L(\Theta_\nu)}
    {4\pi l - 2(\delta + \delta_{\rm th})} \,.
    \label{eq:intphase}
\ee
{}From the comparison of eqs.~(\ref{eq:intphase}) and (\ref{eq:7}) it 
follows that the interference phase curves (shown in cyan in the oscillogram 
of Fig.~\ref{fig:2}) are similar to atmospheric curves, but are steeper than 
the latter for $\nu$'s and less steep for $\bar{\nu}$'s.

\section{CONCLUSIONS}
It can be seen from Fig.~\ref{fig:2} that the three grids of curves 
discussed above -- the solar and atmospheric ``magic'' ones and the 
interference phase curves -- give a very accurate description of the borders 
between the domains. On these borders the probability difference $\Delta 
P_{\mu e}^{\rm CP}=P_{\mu e}(\delta)-P_{\mu e}(\delta_{\rm th})$  
vanishes, and therefore there is no sensitivity to the 
phase $\delta$. The regions of maximal sensitivity to the CP-violating 
phase correspond to the central parts of the domains.

Our analysis has been performed in terms of the oscillation probabilities; 
a more realistic study which considers the number of events in future 
detectors is currently under way.


\end{document}